\documentclass[preprint,superscriptaddress,longbibliography]{revtex4-2}
\usepackage[utf8]{inputenc}
\usepackage[T1]{fontenc}
\usepackage{amsmath}
\usepackage{amssymb}
\usepackage{graphicx}
\usepackage[capitalize, nameinlink]{cleveref}
\usepackage{subfigure}
\graphicspath{{figures/}}
\usepackage{adjustbox}
\usepackage{booktabs}
\usepackage{multirow}

\usepackage{color}
\usepackage{rotating}
\usepackage{ulem}

\begin{document}

\title{Deep learning statistical defect models on magnetic material dynamic and static properties}

\author{C. Eagan}
\affiliation{Center for Magnetism and Magnetic Nanostructures, University of Colorado Colorado Springs, Colorado Springs, CO, USA}
\author{M. Copus}
\affiliation{Center for Magnetism and Magnetic Nanostructures, University of Colorado Colorado Springs, Colorado Springs, CO, USA}
\author{E. Iacocca}
\email{eiacocca@uccs.edu}
\affiliation{Center for Magnetism and Magnetic Nanostructures, University of Colorado Colorado Springs, Colorado Springs, CO, USA}


\begin{abstract}
The modeling of realistic magnetic materials requires the inclusion of defects. Based on the pseudospectral Landau-Lifshitz description of magnetisation dynamics, we propose a statistical model that takes into account defects, specifically vacancies. This statistical model can be integrated with deep learning techniques that correlate defect thresholds with relevant physical observables. We develop a convolutional neural network and a physics-informed neural network combined with theory of functional connections to predict the dispersion relation given defect parameters and physical constraints. A two-branch convolutional neural network is developed to predict domain-wall widths depending on defects threshold, taking into account the spatial profile and domain-wall width separately to achieve a prediction. The proposed physics-informed approaches leverage deep-learning and achieve statistical predictions measured in physical units. This is a stepping stone towards the discovery of new materials and the determination of minimal defect thresholds required for desired dynamics, states, or topological textures.
\end{abstract}

\maketitle
\newpage

Deep learning is ubiquitous in perhaps every aspect of daily life, and is an increasingly common tool in the sciences; fields from biology to physics are increasingly exploring its usefulness to expand on computational techniques~\cite{lecun2015deep}. Machine learning techniques have stepped beyond simple predictions and have now endeavored to harness the power of artificial intelligence for predicting states and desired outcomes~\cite{Goodfellow-et-al-2016}. Applications range from predicting new chemical compounds~\cite{Elton2019}, biological population growth~\cite{Greener2021}, and material sciences~\cite{Zhong2022}. While neural networks are inspired by mammalian brain neural networks, the idea of deep learning is not limited to neural networks, but can rather describe a layered learning process.

Within physics, an alluring concept is the use of physics-informed neural networks~\cite{Raissi2019,Karniadakis2021}, which can predict individual states in well-behaved systems governed by partial differential equations (PDE). Alternatively, the physics described by the PDE can be learned in an abstracted Fourier space using the Fourier neural operator (FNO)~\cite{Li2020b,Li2022} which has been shown to predict the dynamics of two-dimensional flows. However, for non-linear systems governed by more complex PDEs, it can be challenging to predict states or outcomes with a high threshold of certainty with these methods.

So far, the field of material science has been quite successful in using machine learning techniques as well as championing explainability to ``open the black box'', e.g. see Ref.~\cite{Zhong2022, choudhary2022recent} and references therein. In the field of magnetism, machine learning techniques have been used rather sporadically, e.g., to determine their electronic structure~\cite{Nelson2019,Domina2022}, electron-induced torques~\cite{Zhang2023,Tyberg2025}, nucleation fields~\cite{Gusenbauer2020,Bhandari2024}, as well as accelerating magnetisation dynamic simulations~\cite{Schaffer2022,Exl2020,Exl2021,Cai2024}. Magnetic materials have also been used as integral part of potential devices relying on machine learning, including spin-wave based logic gates~\cite{Papp2021,Wang2021} and reservoir computing~\cite{Gartside2022,Stenning2024,Lee2024,Abbott2025}. Another avenue where machine learning has been used is to discover new magnetic materials. Examples of these efforts include the prediction of materials with high coercivity~\cite{Kovacs2023,Halder2025} and magnetic ordering via neural networks~\cite{Xu2025}. However, the impact of defects in realistic materials has remained virtually unexplored. This issue is essential to understand whether the predicted materials are feasible to produce and if their predicted properties are robust to fluctuations. To tackle this problem, we leverage a statistical approach of magnetisation dynamics that takes into account material defects.

In this work, we show how a statistical interpretation of the magnetic nonlocal interactions can be used to consider material defects and to build a dataset aimed at predicting defect thresholds in which topological textures can be stabilized. Our approach is based on the pseudospectral Landau-Lifshitz (PS-LL) equation~\cite{Rockwell2024,Copus2025,Roxburgh2025b,Foglia2024} in which atomic-scale exchange interactions are modeled as a convolution kernel. The model is modified to consider a statistical defect representation, a statistic pseudospectral Landau-Lifshitz (SPS-LL) model, to ensure compatibility with a deep learning methodology. While the approach is completely general, we demonstrate here its applicability within a one-dimensional (1D) spin chain.

The defects are considered statistically as random telegraph noise characterized by parameters associated to defect size and density. Therefore, machine learning approaches can link such quantities to magnetic physical quantities such as the magnon dispersion relation and domain-wall width. To achieve this connection, we develop deep learning approaches targeted at magnetic materials and extendable to material science in general. On the one hand, the magnon dispersion relation is learned by a combination of convolutional neural networks and physics-informed neural networks using theory of functional connections that allow us to constrain the learned parameters to the underlying physical form of the dispersion relation. On the other hand, a convolutional neural network is used to learn the correlation between the magnetisation's domain-wall profile and domain-wall width to the defect parameters.

The approaches presented here can be extended to higher spatial dimensions to stabilize and predict unique topological features in magnetic materials. Recent advances in fabrication and imaging techniques~\cite{Donnelly2015,Donnelly2017,Donnelly2021,Kent2021,Rana2023,DiPietro2023,Girardi2024,Raftrey2024} as well as the recent observation of magnetic hopfions~\cite{Wang2019,Liu2020,Raftrey2021,Kent2021,Balakrishnan2023,Zhang2023,Zheng2023} suggest future progress aided by machine learning approaches and suitable underlying physical models. Our proposed approach to predict defect thresholds can be used for current materials and could be extended to discover new materials based on desired features, for example, specific dynamics or topological textures.

\section*{Results}
\subsection*{Physical model of material defects}

We begin our description with the Landau-Lifshitz equation for magnetisation dynamics, given by
\begin{equation}
\label{eq:ll}
    \frac{\partial}{\partial t}\textbf{m}=-\gamma \mu_0\left[\textbf{m} \times \textbf{H}_\mathrm{eff} + \alpha\textbf{m}\times (\textbf{m} \times \textbf{H}_\mathrm{eff})\right],
\end{equation}
where \textbf{m} denotes the magnetisation vector normalised by the saturation magnetisation, $M_s$, $\gamma$ is the gyromagnetic ratio, $\mu_0$ is the vacuum permeability, $\alpha$ is the Gilbert damping constant used in the Landau-Lifshitz form, valid if $\alpha\ll 1$, and the effective field $\textbf{H}_\mathrm{eff}$ is given by
\begin{equation}
    \mathbf{H}_\mathrm{eff} = \mathbf{H}_l-\mathcal{F}^{-1}\{{\kappa(\mathbf{k})}\hat{\mathbf{m}}\},
\end{equation}
where $\mathbf{H}_l$ represent local fields, such as anisotropy and an external field, and the nonlocal terms are captured by the term $\mathcal{F}^{-1}\{{\kappa(k)}\hat{\mathbf{m}}\}$, where $\kappa(\mathbf{k})$ is a convolution kernel in units of A~m$^{-1}$.

\begin{figure}
    \centering
    \includegraphics[width=5.5in]{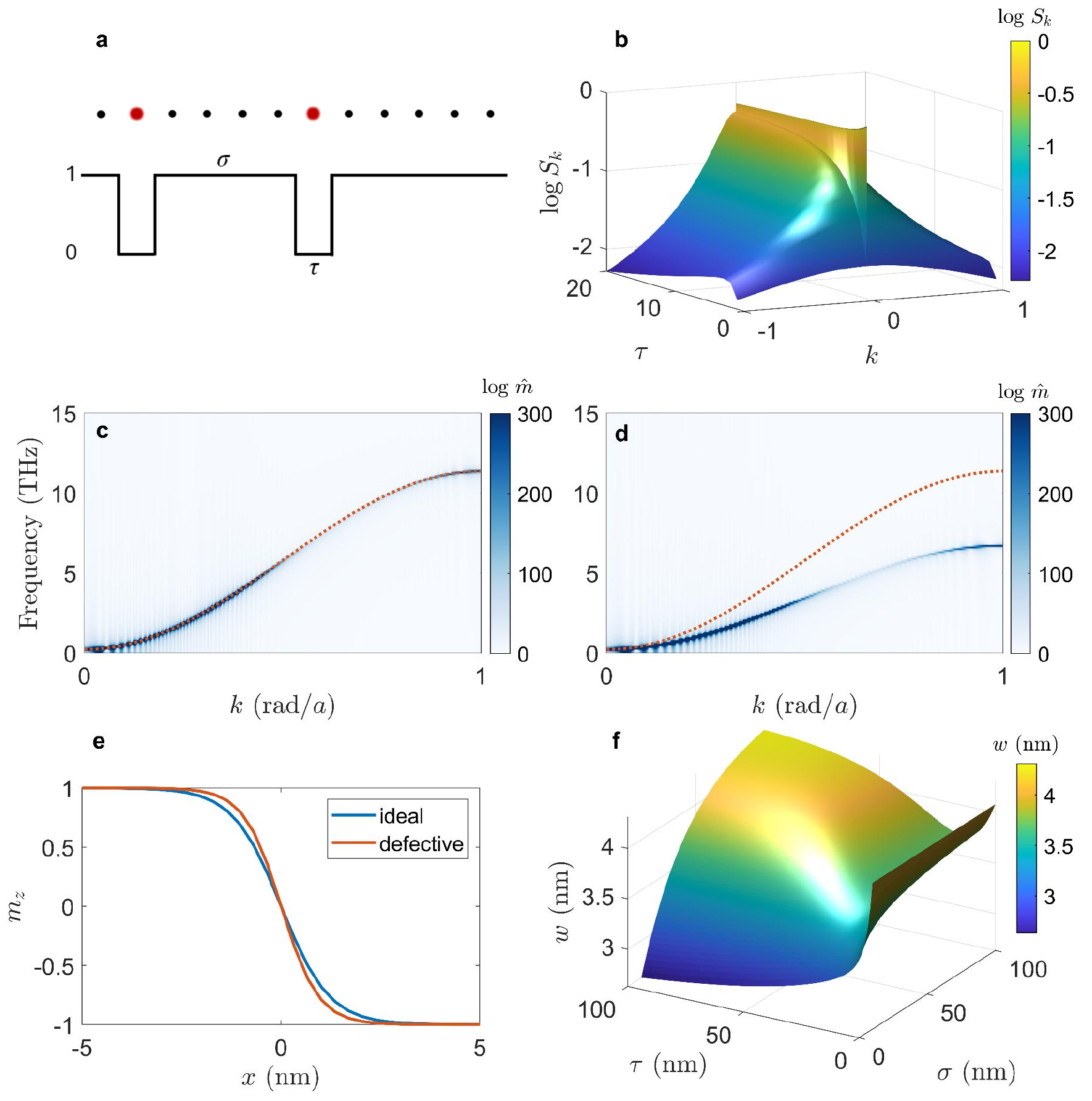}
    \caption{\footnotesize\textbf{Modeling of a defective material.} \textbf{a} A 1D-chain of atoms can have vacancies at random locations, illustrated by red circles. This chain is represented by a digital signal where 1 indicates atoms and 0 represents vacancies. $\sigma$ represents the average length of defect-free atomic chain and $\tau$ represents the average length of vacancies. \textbf{b} The Fourier transform of the RTN signal, Eq.~\eqref{eq:machlup}, shown as a function of $k$ and $\tau$ for $\sigma=10$~nm. When $\tau=0$, $S(k) = \delta(k)$. As $\tau$ increases, the broadband Lorentzian spectrum grows in amplitude. \textbf{c} Dispersion relation calculated for an ideal material and \textbf{d} a material in which $\sigma=10$~nm and $\tau=5$~nm. In both panels, the ideal magnon dispersion relation of Eq.~\eqref{eq:omega} scaled by $\gamma\mu_0M_s$ is shown by a red dashed curve. The color scheme is found in Ref.~\cite{colorscheme}. \textbf{e} Domain wall profile for an ideal material, blue curve $\tau=0$, and a defective material, red curve $\sigma=10$~nm and $\tau=5$~nm. \textbf{f} Surface plot of the the domain-wall width as a function of $\sigma$ and $\tau$, calculated from numerically relaxed domain walls. }
    \label{fig:defectmodel}
\end{figure}

So far, the convolution kernel has been typically chosen to be the magnon dispersion relation~\cite{Rockwell2024,Foglia2024}, augmented with Dzyaloshinskii-Moriya interaction~\cite{Copus2025}, or modified to describe magnetostatic waves~\cite{Roxburgh2025b}. Here, we consider a 1D chain of atoms where defects are simply vacancies. This is illustrated in Fig.~\ref{fig:defectmodel}\textbf{a}, top panel,  where the atomic sites are black circles and the vacancies are red circles. The presence of defects can be represented by the digital signal shown in the bottom panel, where 1 indicates an atom and 0 a defect. Because the defect location is random, this digital signal is simply random telegraph noise (RTN) where $\sigma$ is the mean distance in the 1 state, or the average size of the ideal material, and $\tau$ is the mean distance in the 0 state, or the mean defect size. Thus, $\tau=0$ represents a perfect material.

In a direct simulation, the RTN can be multiplied directly to the magnetisation vector, so that a vacancy describes a numerical cell without magnetisation~\cite{Roxburgh2025b}. In 1D, this would correspond to a small chain of atoms in which wave scattering would be precluded, as otherwise expected physically in 2D and 3D. Therefore, we interpret the RTN spectrally. The power spectral density of RTN was derived in Ref.~\cite{Machlup1954} as
\begin{equation}
\label{eq:machlup}
    S(k)=\frac{1}{\pi}\frac{\sigma \tau}{(\sigma + \tau)^2}\frac{T}{1 + \left(kT\right)^2}+\left(\frac{\sigma}{\sigma +\tau}\right)^2\delta(k)
\end{equation}
where $T=\left(\sigma^{-1}+\tau^{-1}\right)^{-1}$, is the mean defect spatial density. As mentioned before, a perfect material corresponds to $\tau=0$ in which case, Eq~\eqref{eq:machlup} reduces to a delta function. As $\tau$ increases relative to $\sigma$, the delta function is reduced in magnitude and a Lorentzian lineshape appears. This transition is illustrated in Fig.~\ref{fig:defectmodel}\textbf{b} as a surface plot of $S(k)$ as a function of $\tau$ and assuming $\sigma=10$~nm. In this extreme example, the Lorentzian associated with defects begins as a low-amplitude and wide lineshape. As $\tau$ increases, the magnitude of the Lorentzian follows suit and its width is reduced. We note that in the limit of large $\tau$, $S(k)=0$ as also expected from the definition of the RTN signal. More details on the RTN and its relation to Eq.~\eqref{eq:machlup} is provided in the Supplementary Material~\ref{SI:PSD}.

Because the RTN is in real space, its Fourier representation applies as a convolution in Fourier space. The statistical representation of the magnetisation due to defects can be written as
\begin{equation}
    \label{eq:Wk}
    W(k)=\sqrt{S(k)}*\hat{\textbf{m}}(k),
\end{equation}
where $*$ represents convolution and we take the square root of $S(k)$ to consider its Fourier transform rather than power spectral density. For the case of a perfect material, $S(k) = \delta(k)$, and Eq.~\ref{eq:Wk} reduces to $\hat{\textbf{m}}(k)$, as expected. Finally, the nonlocal kernel takes the form
\begin{equation}
    \label{eq:kernel}
    \kappa(k)\hat{\textbf{m}}=M_s\omega(k)W(k)=M_s\omega(k)\left[\sqrt{S(k)}*\hat{\textbf{m}}(k)\right],
\end{equation}
where the dimensionless 1D magnon dispersion relation is given by
\begin{equation}
    \label{eq:omega}
    \omega(k) = 2\left(\frac{\lambda_\mathrm{ex}}{a}\right)^2\left[1-\cos{(ak)}\right],
\end{equation}
$\lambda_\mathrm{ex}$ is the exchange length, and $a$ is the lattice constant. Equations~\eqref{eq:ll} with Eqs.~\eqref{eq:kernel} and \eqref{eq:omega} compose the statistical pseudospectral Landau-Lifshitz (SPS-LL) model.

Because this representation includes both a multiplication and convolution, there is no numerically preferred method to compute it, i.e., Fourier space or real space. However, the approach is based on the fact that the defects can be represented in Fourier space, eliminating the need of large simulations domains and several simulations to obtain statistics. In addition, the representation is grid-independent, meaning that the impact of defects are considered up to the maximally resolved wavenumber and limited by the first Brillouin zone given a lattice constant $a$. A consequence of this statistical interpretation of defects is that the model cannot describe a single object. For example, a domain wall represented with our statistical approach would exhibit a profile analogous to that of an ensemble of domain walls in a material with a given defect size and density.

The function $S(k)$ can be interpreted as a low-pass filter in Fourier space. For example, if one considers a uniform magnetisation state, $\hat{\mathbf{m}}=\delta(k)$, the kernel would be simply $\kappa(k)=\omega(k)\sqrt{S(k)}$. This means that small features are arrested or scattered by defects while large features are largely unaffected. The filter characteristic can be clearly seen by comparing the small-amplitude dispersion relation obtained from ideal and defective materials, shown in Fig.~\ref{fig:defectmodel}\textbf{c} and \textbf{d}. In panel \textbf{c}, we show the ideal ($\tau=0$) dispersion computed numerically for a material with strong perpendicular magnetic anisotropy (PMA) such as FePt. The material parameters and computation method of the dispersion relation are detailed in the Methods. The magnon dispersion relation of Eq.~\eqref{eq:omega} scaled by $\gamma\mu_0Ms$ is shown by a red dashed curve, reflecting that the calculation reproduces the dispersion relation almost exactly. In panel \textbf{d} we use $\sigma=10$~nm and $\tau=5$~nm, which is a rather extreme set of conditions. The numerically calculated dispersion exhibits a reduced frequency as the wavenumber approaches the FBZ, confirming that $S(k)$ acts similar to a low-pass filter.

\subsection*{Domain-walls within the SPS-LL}

Changes in the dispersion relation indicate energy variations of the exchange interaction. Based on the interpretation of $S(k)$ as a low-pass filter, the exchange energy due to defects is lower for small magnetic features. This has direct consequences for solitons, whose profile depends on the balance between anisotropy and exchange, i.e., between nonlinearity and dispersion~\cite{Kosevich1990}. For the 1D PMA ferromagnetic chain considered here, a soliton takes the form of a kink, which is a domain wall~\cite{Braun2012}. The decreasing exchange energy for small features suggests that as the defect density increases, the domain walls become narrower as the anisotropy energy becomes dominant.

To verify the above statement, we stabilize domain walls in our 1D ferromagnetic chain with varying $\sigma$ and $\tau$ parameters. Because the model is 1D, there is no clear preference between a Bloch or N\'{e}el wall~\cite{Hubert2009}. However, Bloch walls are favored in PMA nanowires, so we will assume this profile as the initial condition. In Fig.~\ref{fig:defectmodel}\textbf{e}, we show a comparison between domain walls in an ideal (blue curve) and defective (red curve) material. The domain-wall in the defective material in this case is narrower. This solution should be interpreted as an average domain wall profile. Indeed, in regions where the defect is not present, the material is effectively ideal.

From the stabilized domain walls, their domain-wall width, $w$, can be numerically extracted as specified in the Methods. The width dependence as a function of $\sigma$ and $\tau$ is shown in Fig.~\ref{fig:defectmodel}\textbf{f}. The equilibrium domain-wall width is $4.08\pm0.12$~nm, equivalent to 12 atomic sites, in good agreement with the analytical domain-wall width of $\pi\sqrt{A/K_u}=3.73$~nm. For low values of $\tau$, the domain-wall width is quickly reduced in all cases. After $\tau\approx13$~nm, the domain-wall width tends to decrease for $\tau>\sigma$ and increase for $\tau<\sigma$. These trends agree with the general expectation that as $\tau\rightarrow\infty$, $S(k)\rightarrow0$ so that exchange interactions are disabled, while $\sigma\rightarrow\infty$ leads to $S(k)\rightarrow\delta(k)$ leading to the ideal domain-wall width. The physically relevant case is only when $\tau<\sigma$, meaning that most of the material is present.

\subsection*{Learning defect thresholds}

The overall rationale of our approach is to learn physical features of the material as a function of $\sigma$ and $\tau$ pairs. The features of interest are the dispersion relation as it captures the non-local exchange interactions in the material, and the domain-wall width as an example of a physical observable. For example, given a certain domain-wall width, the neural network will predict the $\sigma$ and $\tau$ pairs with good accuracy, which in turn can be used to compute the dispersion relation. Ultimately, the proposed approach aims to learn quantities that can be measured and controlled experimentally rather than predicting specific states.

\subsection*{Prediction of dispersion relations}

The computed dispersion relations are essentially 2D matrices where each wavenumber and frequency pair has information on amplitude, c.f. Fig.~\ref{fig:defectmodel}\textbf{b} and \textbf{c}. In other words, the data can be assumed to be an image. This suggests the use of convolutional neural networks (CNNs) for the initial recovery of $\sigma$ and $\tau$ from specific defective dispersions.

The CNN works by specifically using ``pooled'', or convolutional layers, which reduce the dimensionality of information in each layer for learning by transforming matrices to tensors~\cite{Li9451544}. This type of neural network detects patterns and edges in the data, making it ideal for learning and predicting visual patterns~\cite{Li9451544}; we designed it to solve the inverse problem. In other words, given a dispersion relation, our network predicts the associated $\sigma$ and $\tau$ pairs.

First, we randomly generate $\sigma$ and $\tau$ pairs and use the SPS-LL model to compute the associated dispersion relations, as detailed in the Methods. The data is passed through a fully connected CNN optimized for feature extraction. We used sixteen layers as well as final output block. Details on the implementation are found in the Methods.

Predicting the specific dispersion curve from the defect parameter pairs is a more challenging task due to the combination of a learned function required to obey known physical constraints of the system. It is tempting to use a physics-informed neural network (PINN), a network that leverages loss minimization to converge on a system solution whose boundary and limit conditions are set by physical constraints \cite{cuomo2022scientific}. However, while this method may converge on a solution for a single state, we are actively seeking to predict a function from a given set of parameters within a profile. The theory of functional connections (TFC) is a way to implement a learning constraint based on the underlying physics~\cite{schiassi2021extreme}. By constraining the learning of the PINN, the TFC approximated form can always analytically satisfy the physical constraints of the system, while maintaining a neural network with unconstrained parameters necessary for learning \cite{ schiassi2021extreme, Raissi2019}. Because, in general, neural networks can learn any function, including one that is non-physical for the physical system's characteristics it is learning, it is necessary to employ techniques to enforce the physical constraints within the learning. This was not problematic when learning defect parameters from the dispersion curve, because we learned only those parameters, and not the function itself.

\begin{figure}[t]
\centering
\includegraphics[trim={2.5in 0.7in 3in 0.5in}, clip, width=3.3in]{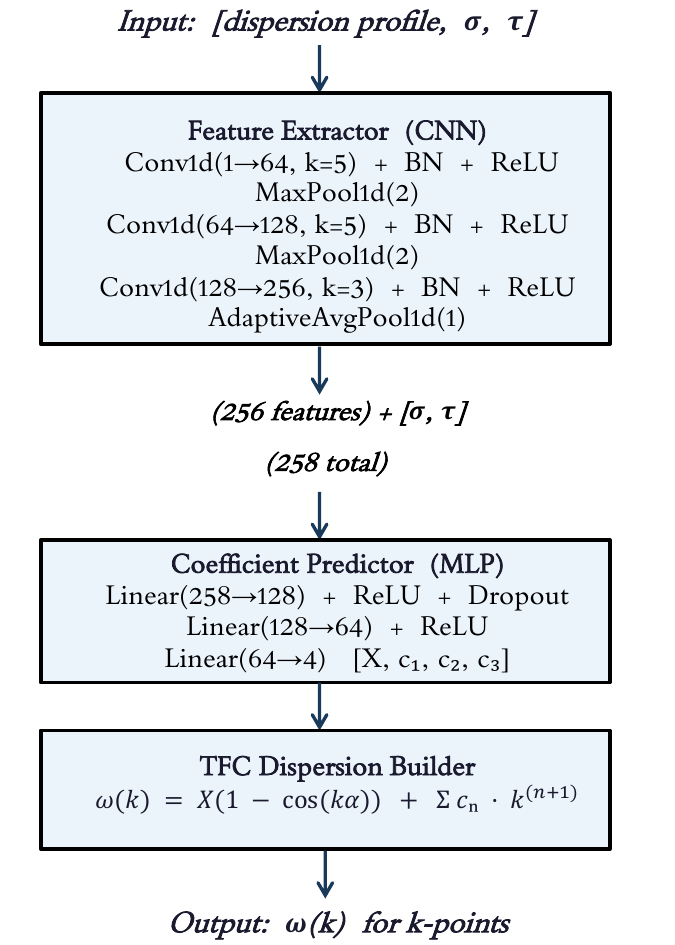}
\caption{\footnotesize\textbf{Neural network architecture for TFC dispersion prediction.} The dispersion relation and $\sigma$ and $\tau$ pairs are given as inputs to a CNN for feature extraction and efficient pooling, with normalization in between each layer. The output is passed to the coefficient predictor that learns based on specific physical parameter constraints and finally a new dispersion is constructed preserving the original structure by the TFC.}
\label{fig:architecture}
\end{figure}

To learn the dispersion function, we combined the prior CNN with a PINN-TFC, for a novel three-block learning approach whose architecture is visualized in Fig.~\ref{fig:architecture}. The CNN block retains the adaptive pooling and extracts features based on the given defect input parameters, then passes these outputs to the second and third PINN-TFC blocks, which learn the appropriate coefficients within given constraints and combine this with learning and predicting the required unchanging functional form of the dispersion \cite{Li9451544, schiassi2021extreme}.

Within the TFC solution, we approach the dispersion prediction as a constrained function optimization learning task. The learning of the specific function $\omega(k)$ is set up to minimize a loss, where the loss is expressed as $L(\omega)$. Utilizing the TFC approach, we can express $\omega(k) = \phi(k,\theta) + \psi(k, \theta)$, where $\phi(k,\theta)$ satisfies  the constraints and contains no learned parameters and $\psi(k, \theta)$ is a trainable term, parametrized during the learning process~\cite{ schiassi2021extreme, Raissi2019}. Therefore, the network does not truly ``know'' physical bounds, but learns the output coefficients, and produces dispersion curves according to defect parameters in accordance with loss minimization.

The constraints are imposed by the initial transformation with the TFC. Thus we predict coefficients that must yield a correct dispersion, bound by statistical material thresholds. We set $\phi(k,\theta)=X(\theta)[1-\cos(k\alpha)]$, where $X$ is a learn coefficient,
\begin{equation}
    \label{eq:TNC}
    X_{learned} \propto \text{Learned Exchange} \in [X]_{\text{material min}}^{\text{material max}},
\end{equation}
and $\psi(k,\omega)=\sum c_n(\theta) k^{n+1}$. Therefore, the TFC-learned dispersion decomposes as follows
\begin{equation}
    \omega(k) = \underbrace{X(\theta)}_{\text{constrained}} \underbrace{[1-\cos(k\alpha)]+\gamma\mu_0H_k}_{\text{fixed physics}} + \underbrace{\sum_{n=1}^5 c_n(\theta) k^{n+1}}_{\text{learned corrections}}.
\end{equation}
Thus, for any given input, we can predict the amplitude and subsequent coefficients according to predicting $(X,c)$ where $X \in [0,5.1]$, constraining the sigmoid functions for each correction with learned $c$ coefficients, which ensures learning within the constraint for $X$ based on the material-specific ideal dispersion. We constrain the coefficient on the torch.sigmoid function in this manner.
\begin{subequations}
\begin{eqnarray}
X_{\text{learned}} &=& 5.1 \times \text{torch.sigmoid}(\text{coeffs}[0, 5.1]), \\
X\hphantom{{}_{\text{learned}}} &=& 5.1 \times \text{sigmoid}(X_{\text{learned}})
\end{eqnarray}
\end{subequations}
\begin{figure}[t]
    \centering
    \includegraphics[width=5in]{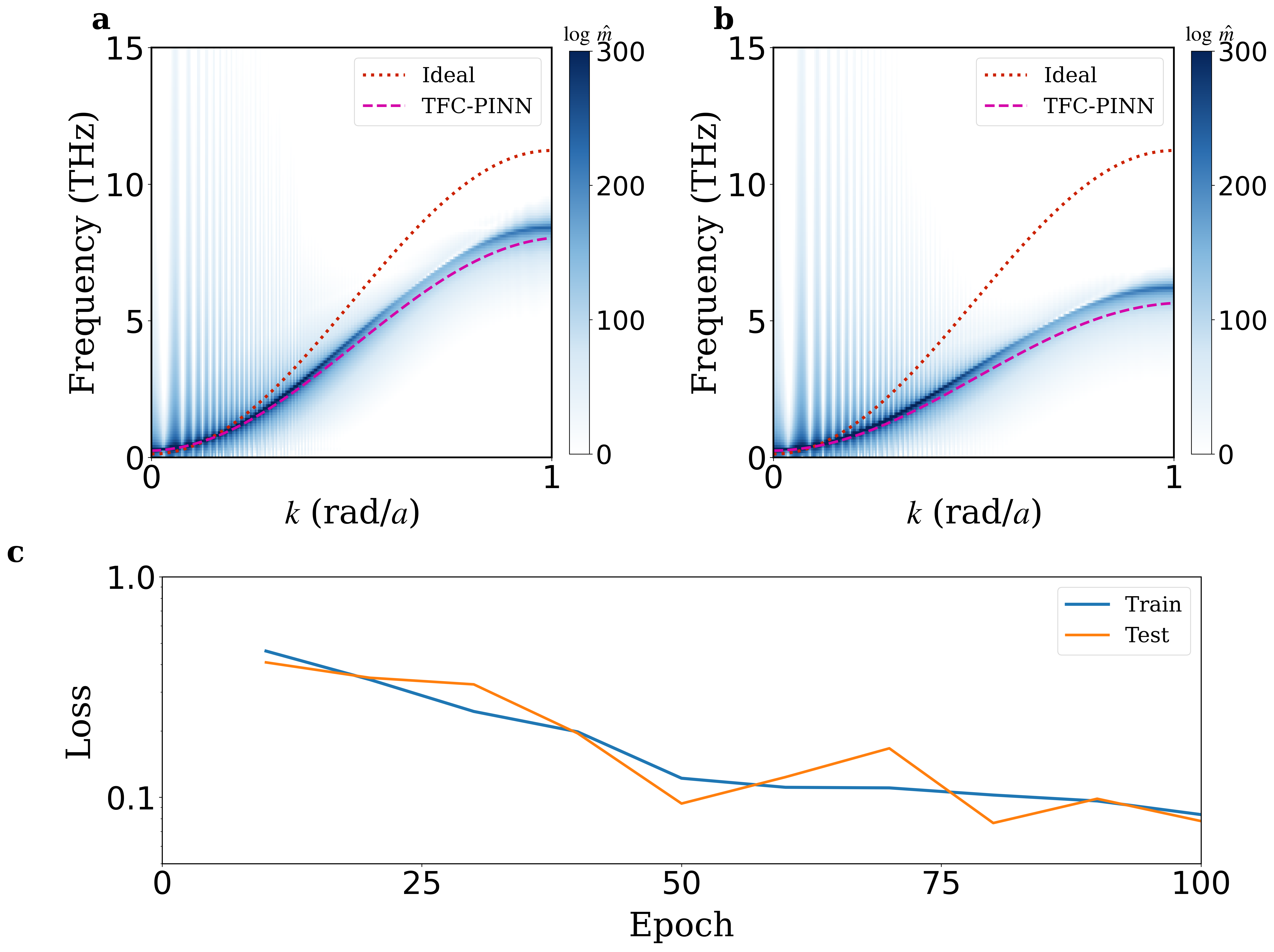}
    \caption{\footnotesize\textbf{Dispersion relation predictions from the TCF-PINN learned function.} Dispersion relations computed by the SPS-LL for \textbf{a} $\sigma = 22.9$~nm, $\tau=3.0$~nm, and $X=3.65$~THz and \textbf{b} $\sigma=15.1$~nm, $\tau=12.4$~nm, and $X=2.45$~THz. The predicted dispersion relation from the given $\sigma$ and $\tau$ are plotted by dashed magenta curves, in good agreement with the numerical solution. The ideal dispersion relation is shown by dotted red curves for reference. The color scheme is found in Ref.~\cite{colorscheme}. \textbf{c} Training loss trend for the TCF-PINN learned function. This training loss plot shows an optimal learning trend.}
    \label{fig:TNC}
\end{figure}

A mean absolute error (MAE) consistently between 0.29 and 0.36 across $k$-values shows an excellent fit on the lower end of a target baseline of MAE between 0.2 and 0.5. As an example, the learned dispersion relations are shown by dashed magenta curves in Fig.~\ref{fig:TNC}\textbf{a} and \textbf{b}. We find a good agreement with the dispersion relation computed by the SPS-LL, shown as a colormap. The error in the prediction stems mostly from the large $k$ components, demonstrating that the numerical dispersion is not exactly a cosine function. This also implies that the polynomial correction is insufficient to fully describe the dispersion. The error could be further reduced by using other correction functions or a combination of functions. Despite this systematic error, the learned dispersion is remarkably close to the numerical dispersion relation. For reference, the ideal dispersion relation is shown by dotted red curves. The evolution of the loss as a function of epochs in shown in Fig.~\ref{fig:TNC}\textbf{c} for the train and test data. The good agreement between both train and test data demonstrates an appropriate learning trend. The proposed architecture is thus able to learn the dispersion relation and correlate its functional form to the $\sigma$ and $\tau$ pairs.

\subsection*{Prediction of domain-wall widths}

A regime of interest is to identify the features of magnetic textures due to material defects. In our current 1D SPS-LL implementation, a natural configuration to investigate is a domain wall. To demonstrate the feasibility of leveraging these deep learning techniques for such an application, we use the initial CNN model that learned the $\sigma,\tau$ pairs from the dispersion relation to predict a domain wall widths. 

\begin{figure}[t]
\centering
\includegraphics[trim={2in .8in 2in 0.8in},clip, width=3.5in]{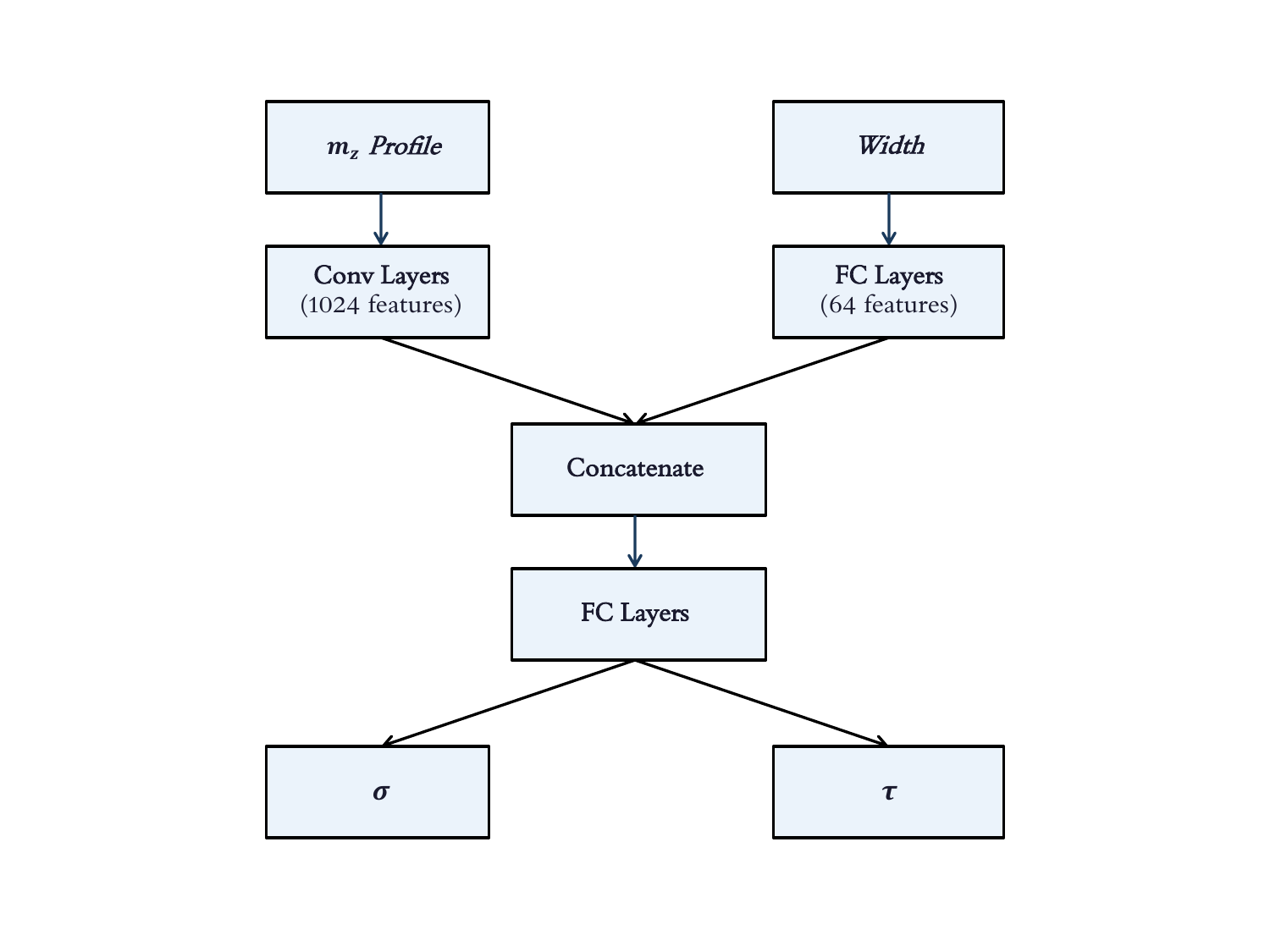}
\caption{\footnotesize\label{fig:architecture2} \textbf{Architecture to predict domain-wall widths.} The CNN takes the magnetization profile and associated domain wall width calculated with the defect noise parameters and returns the defect parameters for that width.}
\end{figure}

To build the database, we stabilize 12,000 Bloch-type domain walls with random $\sigma$ and $\tau$ pairs and estimate their widths, as discussed above and detailed in the Methods. As shown in Fig.~\ref{fig:defectmodel}\textbf{e}, the domain-wall width is multivalued when considering a wide range of $\sigma$ and $\tau$ pairs. For this reason, we constrained our database to physically relevant cases in which the defects are relatively small. This is important for FePt since the expected domain-wall width is comparable to only a few atoms. Therefore, we used $0\leq\sigma\leq100$ and $0\leq\tau\leq 20$. Each domain-wall width is associated to a specific $\sigma$ and $\tau$ pair that is used by a simple CNN to learn their relationship. The input is the $m_z$ profile and the associated domain-wall width, with the output being the predicted $\sigma$ and $\tau$. The CNN architecture is illustrated in Fig.~\ref{fig:architecture2}. Our model uses a dual-branch architecture that processes spatial features separately from the geometric width feature, then fuses them for final prediction. The first layer is a convolutional feature extractor, which processes the $m_z$ profile through 4 convolutional blocks to capture multi-scale spatial patterns.

\begin{figure}[h]
    \centering
    \includegraphics[width=6in]{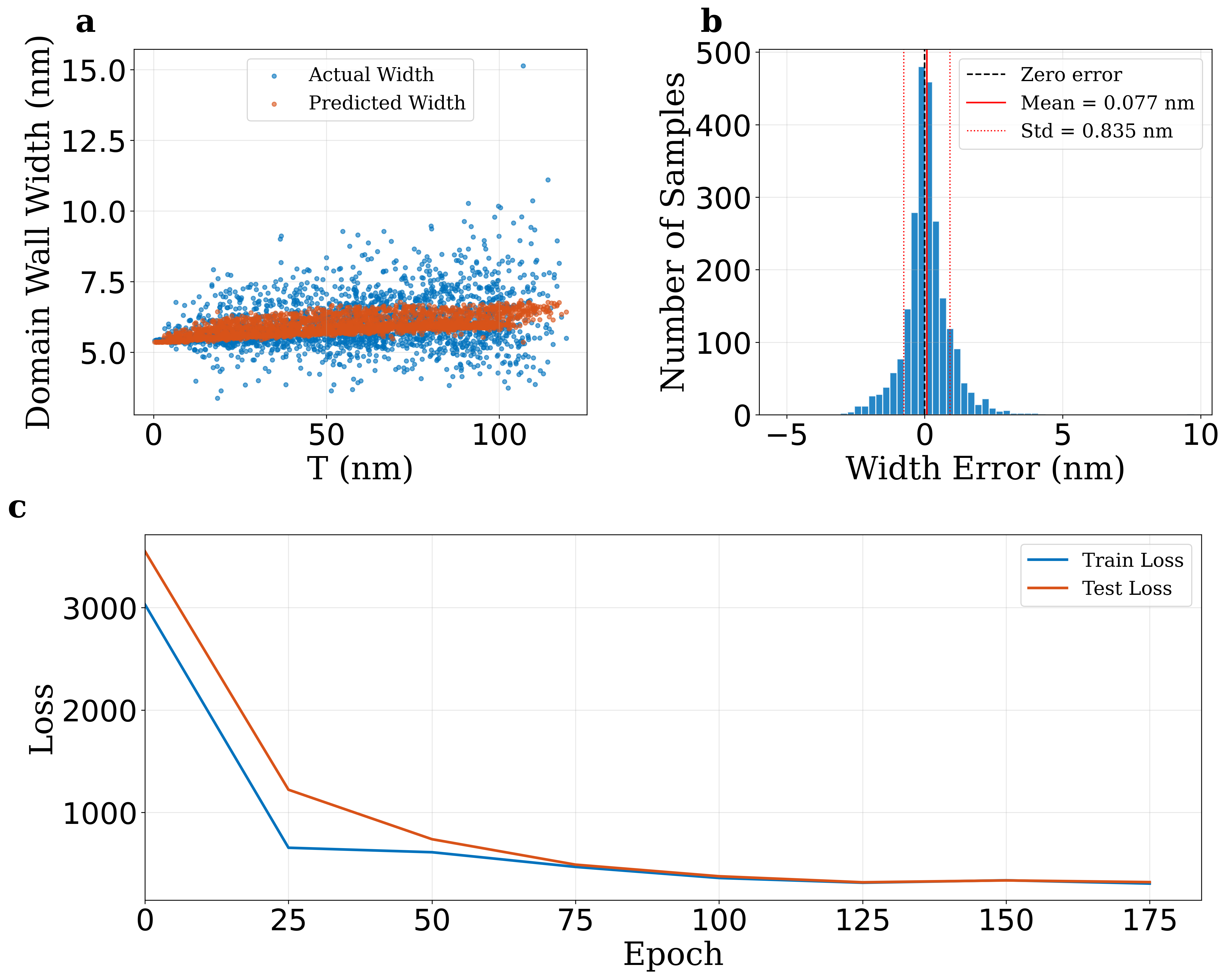}
		\caption{\footnotesize\textbf{Learned domain-wall widths as a function of $\sigma$ and $\tau$}. \textbf{a} Actual (blue symbols) and predicted (red symbols) widths as a function of defect density T (nm). The learned domain-wall widths closely follow the trend expected from numerical calculations. \textbf{b} Histogram of the domain-wall width errors. The mean of the distribution is 0.077~nm and the standard deviation is $0.835$~nm. \textbf{c} Training loss trend for the CNN learned width. For the training we used large batch training in increments of 25 epochs.}
    \label{fig:DWlearning}
\end{figure}

Due to the calculated width from the spectral profile including the defect parameters, overfitting was a challenge. For this reason, we include overfitting detection. Despite these challenges, our model performed reasonably well. In Fig.~\ref{fig:DWlearning}\textbf{a}, we show the actual and predicted domain-wall widths as a function of defect density $T$. Because the same $T$ can be achieved by several combinations of $\sigma$ and $\tau$, there is a significant scatter in the domain-wall widths computed from the SPS-LL-generated profiles, shown in blue circles. The predicted domain-wall width is shown in red circles and closely follows the expected trend.

To gain more insights into the performance of our architecture, Fig.~\ref{fig:DWlearning}\textbf{b} shows the distribution of domain-wall widths errors, computed as the difference between the calculated and learned domain-wall widths. The peak of the distribution is at $0.065$~nm which is under the lattice constant. While there are outliers as expected in statistical learning, the standard deviation of the domain-wall prediction is $0.835$~nm. The training-loss curve using large batch processing is shown in Fig.~\ref{fig:DWlearning}\textbf{c}.

\section{Discussion}

We have demonstrated an approach that leverages a statistical physical model for magnetic materials and deep learning to correlate physical features with material defects. Such a blended technique provides two levels of computational efficiency. First, a statistical model of magnetisation dynamics alleviates the requirement of using large-scale and repeated simulations to accumulate statistics. Second, the use of deep learning and training models enables an accurate and computationally efficient determination of physical observables. In the 1D structure investigated here, these statement are admittedly difficult to quantify because exchange-dominated scattering from defects reduces to simple reflections. Further investigations in 2D or 3D are needed to fully assert the accuracy of the SPS-LL.

Extensions into 2D and 3D are certainly feasible. The PS-LL model has been already demonstrated in 2D in good agreement with other simulation tools~\cite{Copus2025}, known magnetostatic regimes~\cite{Roxburgh2025b}, and ultrafast experiments~\cite{Foglia2024}. A 3D implementation will be presented elsewhere. Utilizing this approach at higher dimensions has the potential to reduce the computational cost even further once a model is trained. In addition, 2D and 3D systems are interesting from a technological point of view because they can host magnetic solitons such as skyrmions and hopfions. In particular, hopfions are receiving significant attention as objects available in materials with bulk Dzyaloshinskii-Moriya interaction or biquadradic exchange interaction. For practical applications, it is important to understand the effect of defects in these textures as well as determining the parameter range in which their existence is guaranteed. Our statistical approach could be useful from these perspectives. For example, a fully trained model could aid in the prediction of material parameters and novel materials hosting and sustaining hopfions as well as providing insights into the minimum material quality.

Finally, our presented model can be also used inversely to predict defect size and densities based on observed data. For example, the dispersion relation can be computed by several methods or even experimentally determined, e.g., by neutron scattering. With this data, the trained model would return the defect size and densities, based on the expectations of the SPS-LL model. Similarly, average domain-wall widths measured by MFM or x-ray scattering could be used as inputs to predict the defect properties. Such an approach would require careful comparison with experiments to assess its accuracy and validity.

In summary, this paper proposes a novel approach to investigate material defects and statistical properties in magnetic materials by taking advantage of deep learning techniques while being firmly rooted in the underlying physics of the problem. As such, we expect our results to be valuable in material science, both in analyzing data and predicting new materials.

\section*{Methods}

\subsection*{Computation of dispersion relation for FePt}

We consider magnetic parameters consistent with FePt in the L1$_0$ phase~\cite{Becker2014}: $M_s=950$~kA~m$^{-1}$, $A=5.1$~pJ/m leading to $\lambda_\mathrm{ex}\approx3.1$~nm, $H_k=7$~MA/m, lattice constant $a=3.39$~\AA{}.

To compute the dispersion relation, we use a single and narrow Gaussian initial condition, as done previously~\cite{Rockwell2024,Copus2025,Roxburgh2025b}. The Gaussian is defined in the $m_x$ magnetisation component for the PMA material implying that the equilibrium direction is zero and the Gaussian is a deviation. We set $m_y=0$ so $m_z=\sqrt{1-m_x^2}$. The standard deviation of the Gaussian function is set equal to the lattice constant so that the initial state resembles a delta function. To excite waves with sufficient amplitude, we set a low Gilbert damping constant of $\alpha=0.001$. In other words, this low damping allow us to obtain a larger amplitude in Fourier space and thus clearly identify the eigenfrequencies. The simulation domain is set to 500 cells discretized at the lattice constant while the simulation runs for 10~ps and is sampled at 10~fs. The space and time evolution of the $m_z$ component is recorded. Therefore, the dispersion relation is obtained with a two-dimensional fast Fourier transform. Given the above spatiotemporal parameters, we obtain a spatial wavenumber resolution of $37$~rad~$\mu$m$^{-1}$ and for the FBZ of $9.26$~rad~nm and a frequency resolution of 100~GHz with a maximum frequency of 50~THz.

\subsection*{Domain-wall simulation and determination of their width}

Domain walls are simulated by initializing the magnetisation with the anzats $m_z=\tanh{(\sqrt{K_U/A_\mathrm{ex}}x)}$, $m_y=0$ and $m_x=\sqrt{1-m_z^2}$, where $K_U=\mu_0M_s(H_k-M_s)/2$. To accelerate convergence, we set $\alpha=1$ and run the simulation for 100~ps.

Once the domain-wall is stabilised, we compute the domain-wall width by fitting its shape with a sum of basis functions. First, we duplicate and flip the $m_z$ magnetisation component to define a domain. In this way, the function is even and we can use a cosine series as a basis function. Then, we compute the polar angle as $\theta=\arccos{(m_z)}$ and compute the coefficients for each harmonic in order to obtain an approximation for the domain, $f(x)$. We then take the first-order Taylor approximation of our basis to obtain the linear limit at the position of one domain wall, for example $L/4$ where $L=1000$ is the duplicated length of the simulation. This means that the linear approximation of the domain wall is
\begin{equation}
    \label{eq:lineardomainwallwidth}
    f(x)=\frac{\pi}{4}+\sum_{n=1}^\infty{a_n\cos{\left(\frac{n\pi}{4}\right)}-\frac{n\pi}{L}\sin{\left(\frac{n\pi}{4}\right)}\left(x-\frac{L}{4}\right)}
\end{equation}
Finally, we define the domain-wall width from the bounds of the linear limit between $0$ and $\pi$. Because the bounds must be obtained from the sum, we truncate the approximation at $n=700$ and then fit the linear approximation with a line so that the domain-wall width can be obtained exactly. This approach corresponds to the definition of domain-wall width as $\pi\sqrt{A_\mathrm{ex}/K_U}$. The error in the determination of the domain-wall comes from the truncation of the sum. In this spirit, the error is estimated as the standard deviation from the last 100 harmonics.

\subsection*{CNN for dispersion relation feature extraction}

The input for the CNN takes the dispersion relation, estimated as the maximum of each $k$-dependent spectrum. This results in 1D dispersion relations that are represented as arrays of length 500 and then used as the CNN's channel input. We then leveraged sixteen total fully connected convolutional layers for feature extraction, organized into three dense blocks and the final output block. Layer 1 has 512 neurons, layer 2 goes from 512 to 256 neurons, layer 3 goes from 256 neurons to 64, and the final output layer goes from 64 back down to our 2 desired outputs ($\sigma$ and $\tau$). The initial layers detect local patterns in the dispersion relation, and the deeper layers learn how these patterns relate to the statistical defect parameters. The convolutional layers learn which spatial characteristics of the dispersion (width, amplitude, gradients) correlate with specific $\sigma,\tau$ combinations. The dense layers perform the final mapping from spatial features to the two defect parameters to be predicted.

The first group of convolutional layers (or first block) is for 1 to 64 channels, with a $1\times5$ kernel, which extracts the basic local features from the dispersion pattern. We use MaxPool which reduces the spatial dimension by a factor 2 (with each block) in the first three blocks. The second block, channels 64 to 128, kernel =5 captures any complex spatial relationships. The third convolutional block channels 128 to 256, kernel =3 learns the higher level abstraction features, from 256 to 128, with kernel=3. 

We use a database of $1000$ simulations. The design features batch normalization with a 30~\% dropout rate to avoid overfitting for a relatively small sample size. We use standard ReLU activation function for non-linearity, standard MSE loss function, an Adam Optimizer, and learning rate stopping to save the best model based on validation performance.

\subsection*{Implementation of TFC}

By using a sigmoid activation function (\texttt{torch.sigmoid}) for $A$, we can ensure to always recover the correct amplitude bounded by the constraints of the perfect dispersion for the specified material. The constraints vary according to the material-specific dispersion. We use a tanh activation function for learned coefficients (\texttt{torch.tanh}). Therefore, by construction,
\begin{enumerate}
    \item $ X_{learned} \in [0, 5.1]: X_{learned} = 5.1\sigma(z)$, where $\sigma(z) \in [0,1]$
    \item $c_n \in [-0.1, 0.1]: c_n = 0.1 \tanh(z) \text{where} \tanh(z) \in [-1,1]$
    \item Combining the learned and unchanging form = $X_{learned}(1-\cos(k\alpha))$
    \item Any corrections are small and bounded by the given physical system constraints.
\end{enumerate}

This means that every network output satisfies the following based on both learned and fixed parameters
$\omega(k) = X_{learned}(1-\cos(k\alpha)) + \epsilon(k)$, where $|\epsilon(k)|$ is bounded and small. Additionally, for small $k$, we still recover the appropriate $k^2$ solution with this method.

\subsection*{Loss function to predict $\sigma,\tau$ from domain walls}

The MSE function is represented to accommodate for the individual defect parameter ranges, as Total Loss = MSE($\sigma$) + 5.0 x MSE($\tau$), with $\tau$ weighted higher in the loss function to compensate for its lower value. The training and evaluation process of the model is structured in the typical epoch progression. Early epochs, 1-50 shows high initial loss due to random weighting. This is followed by mid and late epoch training stages 51-200, with steady loss reduction, and occasional plateaus. Fine-tuning phase with potential learning rate reductions occurred when the ReduceLROnPlateau scheduler activated due to plateaus. These were minimized by a validation loss process, and typical dropout and batch normalization prevent early overfitting.

\section*{Acknowledgments}

This work was supported by the U.S. Department of Energy, Office of Basic Energy Sciences under Award No. DE-SC0024339.

\section*{Author Contributions}

C.E. developed the deep learning algorithms. C.E. and E.I. developed the PS-LL model incorporating random telegraph noise. All authors
contributed with performing validation simulations, analyzing the data, and writing the paper.

\section*{Competing Interests statement}

The authors declare no competing interests. 

\section*{Corresponding authors}
\noindent Correspondence to\\eiacocca@uccs.edu

\bibliographystyle{apsrev4-2}

\onecolumngrid
\clearpage

\textbf{\large{Supplementary material: \textbf{Deep learning statistical defect thresholds effects on magnetic materials dynamic and static properties}}}
\\

C. Eagan, M. Copus, and E. Iacocca
\newpage

\setcounter{section}{0}
\renewcommand{\thesection}{SI \arabic{section}}

\setcounter{figure}{0}
\renewcommand{\thefigure}{SI \arabic{figure}}

\setcounter{table}{0}
\renewcommand{\thetable}{SI \Roman{table}}

\setcounter{equation}{0}
\renewcommand{\theequation}{SI \arabic{equation}}

\section{Power spectral density verification}
\label{SI:PSD}

The function $S(k)$ is a power spectral density and thus has units of m/rad, i.e., inverse wavenumber. However, the definition of the PSD is normalised in amplitude so that the spectral component is aptly scaled to the DC component to 1. This is critical for our approach since the function $S(k)$ must be order one to take its square root and compute the convolution with the magnetisation. The magnetisation is not normalised by the length of the dataset, which eventually leads to a correct inverse fast Fourier transform.

Here, we demonstrate that $S(k)$ indeed corresponds to the normalized PSD of random telegraph noise (RTN). Because the signal is noisy, a direct fast Fourier transform method is not appropriate. Instead, we use the Weiner-Khinchin theorem, which indicates that the Fourier transform of a noisy signal is equal to the Fourier transform of its autocorrelation. DC terms are not noisy, so this component should be treated separately.

\begin{figure}[t]
\begin{center}
\includegraphics[width=6in]{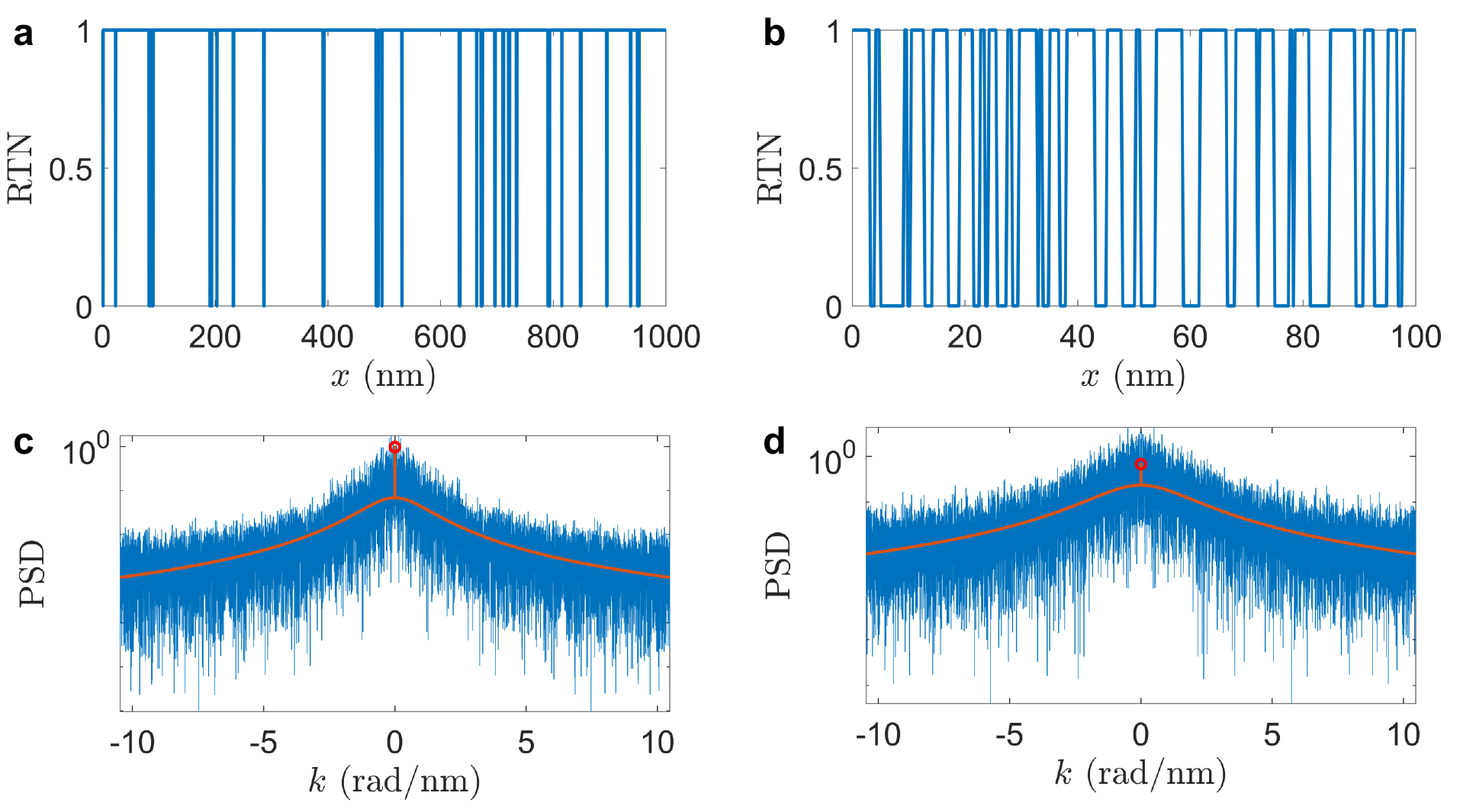}
\caption{\textbf{Computation of RTN spectra.} RTN traces for \textbf{a} $\sigma=100$~nm and $\tau=5$~nm, and \textbf{b} for $\sigma=10$~nm and $\tau=5$~nm. The corresponding PSD computed via Weiner-Khinchin theorem are shown in \textbf{c} and \textbf{d}. The functions $S(k)$ for each case are shown by red curves. }
\label{SI:figRTN}
\end{center}
\end{figure}

First, we generate a RTN by specifying $\sigma$ and $\tau$ and computing the rates per unit length, $r_\sigma=\sigma/dx$ and $r_\tau=\tau/dx$, where $dx$ is the cell size of the $x$ axis. With these normalized rates, we generate exponential random lengths with the \texttt{exprnd} function in MATLAB. These lengths are appended to the array depending on the current value of the RTN. In other words, if the current state is 1, we use $r_\sigma$ and flip the state to 0. If the current state is 0, we use $r_\tau$ and flip the state to 1. Examples of RTNs generated with this method are shown in Fig.~\ref{SI:figRTN}\textbf{a} for $\sigma=100$~nm and $\tau=5$~nm, displayed to $1000$~nm, and in \textbf{b} for $\sigma=10$~nm and $\tau=5$~nm, displayed to $100$~nm. Note that the full RTN signal is computed to $3000$~nm with a cell-size of $dx=0.3$~nm.

The spectrum is computed as follows
\begin{equation}
    \label{eq:si1}
    \mathrm{PSD}(k) = \frac{|\mathcal{F}\{R_{xx}(dx)\}|^2}{L}+\left(\langle\mathrm{RTN}\rangle\right)^2\delta(k),
\end{equation}
where $\mathcal{F}\{\cdot\}$ is a fast Fourier transform, $R_{xx}(dx)$ is the autocorrelation of the RTN, $L$ is the length of the RTN, and $\langle\mathrm{RTN}\rangle$ is the mean of the RTN.

The computed PSD are shown in Fig.~\ref{SI:figRTN}\textbf{c} and \textbf{d} corresponding to the RTN shown in \textbf{a} and \textbf{b}, respectively. In both panels, the function $S(k)$ is shown in red curves, demonstrating good agreement with the computed PSDs. Because the PSDs are already normalized to length (in amplitude), we see that both representations agree. The advantage of using $S(k)$ is that the function considers an infinite signal and so the noise is minimized to a deterministic function. More importantly, the mean is normalized to 1 for an ideal material and tends to zero for no material.

\end{document}